\documentclass[twocolumn,prl,aps,superscriptaddress,showpacs]{revtex4}

\usepackage{bm}\usepackage{graphicx}

\begin{document}

\title{Switching of the magnetic order in CeRhIn$_{5-x}$Sn$_{x}$\\ in the vicinity of its quantum critical point} 

\author{S. Raymond}
\affiliation{Univ. Grenoble Alpes, INAC-SPSMS, F-38000 Grenoble, France}
\affiliation{CEA, INAC-SPSMS, F-38000 Grenoble, France}
\author{J. Buhot\footnote{Present Address : Laboratoire Mat\'eriaux et Ph\'enom\`enes Quantiques, UMR 7162 CNRS, Universit\'e Paris Diderot, 75205 Paris, France.}}
\affiliation{Univ. Grenoble Alpes, INAC-SPSMS, F-38000 Grenoble, France}
\affiliation{CEA, INAC-SPSMS, F-38000 Grenoble, France}
\author{E. Ressouche}
\affiliation{Univ. Grenoble Alpes, INAC-SPSMS, F-38000 Grenoble, France}
\affiliation{CEA, INAC-SPSMS, F-38000 Grenoble, France}
\author{F. Bourdarot}
\affiliation{Univ. Grenoble Alpes, INAC-SPSMS, F-38000 Grenoble, France}
\affiliation{CEA, INAC-SPSMS, F-38000 Grenoble, France}
\author{G. Knebel}
\affiliation{Univ. Grenoble Alpes, INAC-SPSMS, F-38000 Grenoble, France}
\affiliation{CEA, INAC-SPSMS, F-38000 Grenoble, France}
\author{G. Lapertot}
\affiliation{Univ. Grenoble Alpes, INAC-SPSMS, F-38000 Grenoble, France}
\affiliation{CEA, INAC-SPSMS, F-38000 Grenoble, France}

\date{\today}

\begin{abstract}

We report neutron diffraction experiments performed in the tetragonal antiferromagnetic heavy fermion system CeRhIn$_{5-x}$Sn$_{x}$ in its ($x$, $T$) phase diagram up to the vicinity of the critical concentration $x_c$ $\approx$ 0.40, where long range magnetic order is suppressed. The propagation vector of the magnetic structure is found to be $\bf{k_{IC}}$=(1/2, 1/2, $k_l$) with $k_l$ increasing from $k_l$=0.298 to $k_l$=0.410 when $x$ increases from $x$=0 to $x$=0.26. Surprisingly, for $x$=0.30, the order has changed drastically and  a commensurate antiferromagnetism with $\bf{k_{C}}$=(1/2, 1/2, 0) is found. This concentration is located in the proximity of the quantum critical point where superconductivity is expected.

\end{abstract}

\maketitle


The interplay between magnetism and superconductivity is one of the most studied topics in the physics of strongly correlated electron systems. The occurrence of competing or coexisting antiferromagnetic and superconducting ground states is common to many systems: high-$T_c$ cuprates, new iron-based superconductors and heavy fermion (HF) compounds \cite{Uemura}. In this context, the family of HF compounds CeMIn$_{5}$ (M= Co, Rh, Ir), the so-called 1-1-5 compounds, is a fabulous playground since the chemical substitution, the application of pressure or magnetic field lead to the possibility to tune the N\'eel temperature $T_N$ and the superconducting transition temperature $T_c$ to different levels with either $T_N \geq T_c$ or $T_N \leq T_c$ \cite{Sarrao,Knebel}. The parent compound CeRhIn$_{5}$ crystallizes in the tetragonal space group P4/mmm. It orders magnetically in an incommensurate helicoidal structure below 3.8 K at ambient pressure. Pressure induced superconductivity occurs above 1 GPa and at around 2 GPa, the N\'eel temperature equals the superconducting transition temperature. At higher pressure antiferromagnetism is superseded by a pure superconducting state. However a magnetic field, applied in the basal plane of the the tetragonal structure,  inside this superconducting phase, restores an antiferromagnetic order. This phase exists even far above the superconducting upper critical field $H_{c2}$ \cite{KnebelP,Park}. Such a field induced antiferromagnetism bears similarity to the one observed in CeCoIn$_{5}$ out of the purely $d$-wave superconducting state, although in this latter case, the magnetic order disappears at $H_{c2}$ \cite{Kenzelmann}.

Microscopic informations on the magnetic structures are essential in order to grasp the different ingredients at play. In CeRhIn$_{5}$, this is provided essentially by NQR \cite{Yashima} since the triple conditions of high magnetic field, high pressure and low temperature preclude to perform neutron diffraction experiments, which were carried out either under pressure \cite{Raymond_P,Aso} or under magnetic field \cite{Raymond_H}. Another possible route is to substitute Sn for In, which acts as a positive pressure in the the phase diagram. This substitution corresponds to electron doping. In CeRhIn$_{5-x}$Sn$_{x}$, a quantum critical point occurs for $x_c$ $\approx$ 0.40 \cite{Bauer, Donath1, Donath2} and pressure induced superconductivity is reported in CeRhIn$_{4.84}$Sn$_{0.16}$ above 0.8 GPa with however a reduced maximum value of $T_{c}$ \cite{Mendonca}. In the present work,  we determine the evolution of the magnetic structure as a function of $x$ in CeRhIn$_{5-x}$Sn$_{x}$.



Single crystals of  CeRhIn$_{5-x}$Sn$_{x}$ were grown by the self flux method \cite{Canfield} starting with a ratio Ce : Rh : In : Sn = 1 : 1 : 20 : $y$. 
In Ref.\cite{Bauer}, a linear relationship between the actual Sn concentration, $x$, in the crystal and the starting Sn ratio $y$ in the flux
has been found with $x$=0.4$y$. The same relation in the determination of the actual concentration is taken throughout this article since bulk measurements performed on samples of the same batch of the one for the neutron diffraction experiments, which preliminary report can be found in Ref. \cite{Knebel2}, indicate consistent values of $T_{N}$ with the study of Bauer et al. \cite{Bauer}.
Rectangular-shaped samples were cut for neutron scattering experiment for $x$=0.10, 0.16, 0.20, 0.26 and 0.30 with dimensions given in Table I. 

\begin{table}[b!]
\caption{Experimental conditions. The sample size is given along the $a$, $b$ and $c$-axis directions (in this order).}
\begin{ruledtabular}
\begin{tabular}{ccc}
$x$ & Sample size (mm$^{3}$) & Sample Environnement \\
\hline
0.10 & 2$\times$2$\times$1 & $^{4}$He cryostat\\
0.16 & 2.1$\times$1.3$\times$1.8 & $^{3}$He cryostat\\
0.20 & 2$\times$2$\times$1 & $^{3}$He cryostat\\
0.26 & 2$\times$2$\times$2 & $^{3}$He and $^{3}$He-$^{4}$He dilution cryostat\\
0.30 & 3$\times$3$\times$1  & $^{3}$He-$^{4}$He dilution cryostat\\
\end{tabular}
\end{ruledtabular}
\label{table}
\end{table}

The measurements were performed on the two-axis D23-CEA-CRG (Collaborating Research Group) thermal-neutron
diffractometer equipped with a lifting detector at the Institut Laue Langevin, Grenoble.
A copper monochromator provides an unpolarized beam with a wavelength of $\lambda$ = 1.283 $\AA$.
The samples were mounted in different kinds of cryostats accordingly to their respective N\'eel temperatures (See Table I). The [1, -1, 0] direction was set as the vertical axis. 
For each sample, crystal and magnetic structures were refined and details of the method are given in a previous study performed on CeRhIn$_{5}$ \cite{Raymond_H}.
The additional parameters compared to CeRhIn$_{5}$ are the occupations of the two inequivalent In sites by Sn substituent (In(1) in the Ce-plane and In(2) in between these planes). A previous crystallographic study performed using a neutron four circle diffractrometer on $x$=0.16 suggests that Sn preferentially occupies the In(1) site (66 \%) compared to the In(2) site (34 \%) \cite{Knebel2}. Similar conclusion is drawn from the NQR results obtained for $x$=0.044 \cite{Rusz}. In the present study, it is not possible to discuss this point. However this does not affect the results on the magnetic structure determination.


For each concentration, the search for the magnetic propagation vector was made by doing wide scans along $\bf{Q}$=(1/2, 1/2, $l$). In this paper, the scattering vector, $\bf{Q}$, is written as $\bf{Q}$=${\bm \tau}$+$\bf{q}$ where ${\bm \tau}$ is a Brillouin zone center and $\bf{q}$=($h$, $k$, $l$). All coordinates are expressed in reciprocal lattice units (r.l.u.). All the raw data shown in the figures of the paper were collected in the first Brillouin zone where $\bf{Q}$=$\bf{q}$. Representative $\bf{Q}$-scans measured for $x$=0.16 along [0,0,1] and [1,1,0] directions at $T$=0.4 K are shown in Figure 1.
\begin{figure}
\centering
\includegraphics[width=8cm]{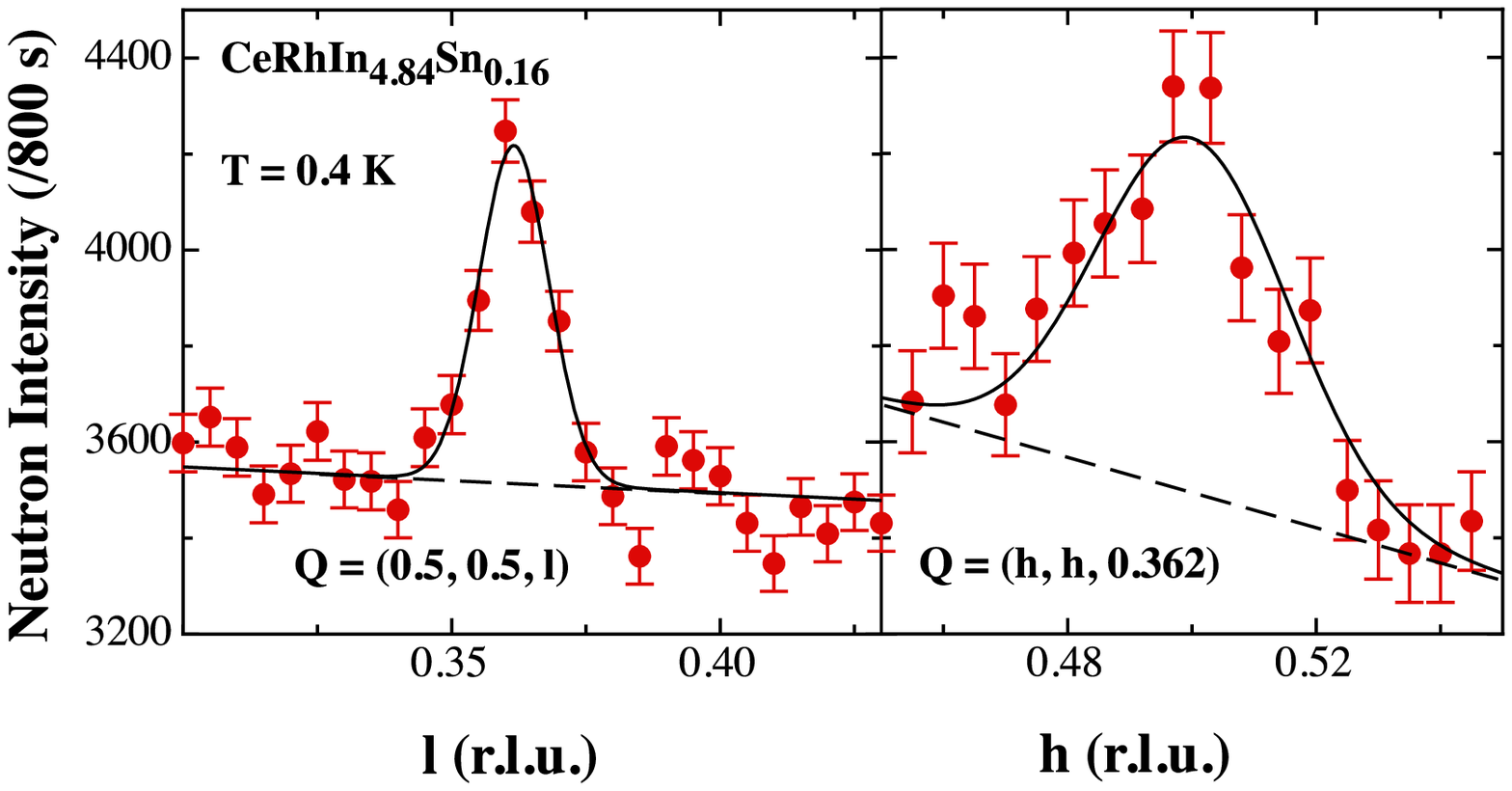}
\caption{$\bf{Q}$-scans performed along the [0,0,1] and [1,1,0] directions for CeRhIn$_{4.84}$Sn$_{0.16}$ at $T$=0.4 K. The  full lines are Gaussian fits and the dash lines indicate the background.}
\end{figure}
Figure 2 shows similar scans performed along $\bf{Q}$=(1/2, 1/2, $l$) with normalized intensities for $x$=0.10, 0.26 and 0.30 for temperatures below and above the respective $T_{N}$ of each sample.
These data show the smooth evolution of the $c$-axis component of the propagation vector between $x$=0.10 and $x$=0.26. For $x$=0.30, the magnetic peak position becomes commensurate with a zero $c$-axis component.
It was also checked for each $x$, that a unique propagation vector exists. This is shown in Fig. 2 for $x$=0.26, where no commensurate signal is evidenced (full circles) and for $x$=0.30 where no incommensurate signal exists (full square). The main result of this paper is the evidence for a switching of incommensurate magnetic order with $\bf{k_{IC}}$=(1/2, 1/2, $k_l$) (0.298 $\leq$ $k_l$ $\leq$ 0.410) to commensurate magnetic order with $\bf{k_{C}}$=(1/2, 1/2, 0) (so-called C-type magnetic structure) in CeRhIn$_{5-x}$Sn$_{x}$ above $x$=0.26.

\begin{figure}
\centering
\hspace{-1.5cm}
\includegraphics[width=8cm]{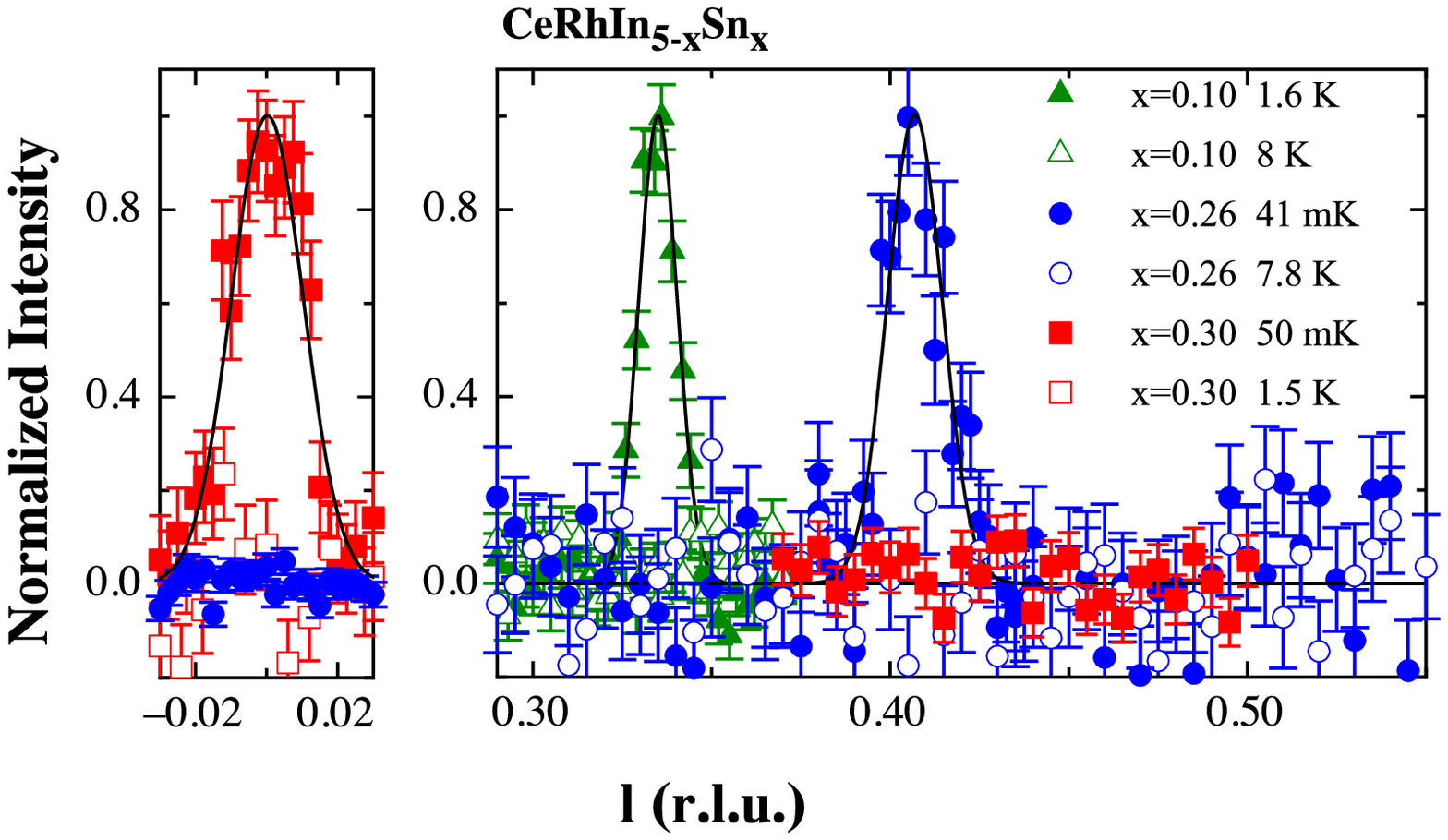}
\caption{$\bf{Q}$-scans performed along the [0,0,1] direction for $x$=0.10, 0.26 and 0.30 for temperatures below and above their respective N\'eel temperatures. The full lines are Gaussian fits.}
\end{figure}

\begin{table}[b!]
\caption{Experimental results for the propagation vector $\bf{k}$, the N\'eel temperature, $T_{N}$, and the ordered moment $M_{0}$.}
\begin{ruledtabular}
\begin{tabular}{cccc}
$x$ & $\bf{k}$ & $T_{N} (K)$ & $M_{0}$ ($\mu_B$)\\
\hline
0 & (0.5, 0.5, 0.298) & 3.80 (1)& 0.59 (2)\\
0.10 & (0.5, 0.5, 0.335) & 3.30 (2)& 0.58 (2) \\
0.16 & (0.5, 0.5, 0.362) & 2.73 (3)& 0.49 (2)\\
0.20 & (0.5, 0.5, 0.389) & 2.03 (3)& 0.59 (4)\\
0.26 & (0.5, 0.5, 0.410) & 1.54 (9)& 0.28 (2)\\
0.30 & (0.5, 0.5, 0) & 0.84 (3)& 0.25 (2)\\
\end{tabular}
\end{ruledtabular}
\label{table}
\end{table}

Figure 3 shows the temperature dependence of a $\bf{Q}$-scan performed along [0,0,1]  for $x$=0.20. The $c$-axis value of the propagation vector, $k_{l}$, does not change significantly with temperature although we cannot exclude a small shift to a lower value in the vicinity of $T_{N}$. The temperature dependence of the order parameter (proportional to the square root of the background subtracted neutron intensity, $I$) was therefore measured on the maximum of the Bragg peak position for each concentration.
Normalized intensities ($I/I_{0}$) are shown in Figure 4.
The N\'eel temperature given in Table II is obtained from a phenomenological description of these curves with $I/I_{0}$ $\propto$ $1-(T/T_N)^{\alpha}$ with $\alpha$ a free parameter. The weakness of the signal does not allow to distinguish between Bragg and diffuse scattering in the vicinity of the phase transition. The best fit is obtained with $\alpha$ $\approx$ 0.6 for $x$=0.10, 0.16, 0.20 and $\alpha$ $\approx$ 0.2 for $x$=0.26, 0.30. 
This change of behavior could originate from different intrinsic magnetic properties near $x_c$, different distributions of concentration or it could be an artifact of this phenomenological method used to determine the N\'eel temperature. 
The $T_{N}$ values obtained are compatible with the one reported by bulk measurements, having in mind that the determinations from specific heat and resistive anomalies show also some differences among themselves \cite{Bauer,Mendonca,Donath1,Donath2}.

Similarly to CeRhIn$_{5}$ \cite{Raymond_H}, the magnetic moment was calculated assuming an helicoidal structure up to $x$=0.26. For $x$=0.30, the commensurate propagation vector implies a collinear structure. The ordered moments are assumed to be in the basal plane of the tetragonal structure as for the pure compound \cite{Raymond_H}. Consequently for $x$=0.30, two magnetic domains corresponding to two orthogonal directions of magnetic moments are considered (an equal domain population is assumed).
The values of the propagation vector, $\bf{k}$, the N\'eel temperature, $T_{N}$, and the ordered moment, $M_{0}$, are summarized in Table II and in Figure 5. The variation of the N\'eel temperature with $x$ is smooth and agrees with bulk measurements.
The magnetic moment evolves only slightly up to $x$=0.20 and then decreases significantly. 
The $c$-axis component of the propagation vector increases linearly with $x$ up to $x$=0.26, following $k_{l}(x)$=0.295(4)+0.44(2)$\times$$x$. For $x$=0.30, the propagation vector has switched to $\bf{k_{C}}$=(1/2, 1/2, 0).
The lines drawn for $T_{N}(x)$ and $M_{0}(x)$ would suggest a critical concentration near 0.35.
This is in agreement with the reported value for $x_c$ that lies between 0.35 \cite{Bauer} and 0.40 \cite{Donath2} depending if an upturn of $T_{N}(x)$ around $x_c$ is considered or not.

\begin{figure}
\centering
\includegraphics[width=8cm]{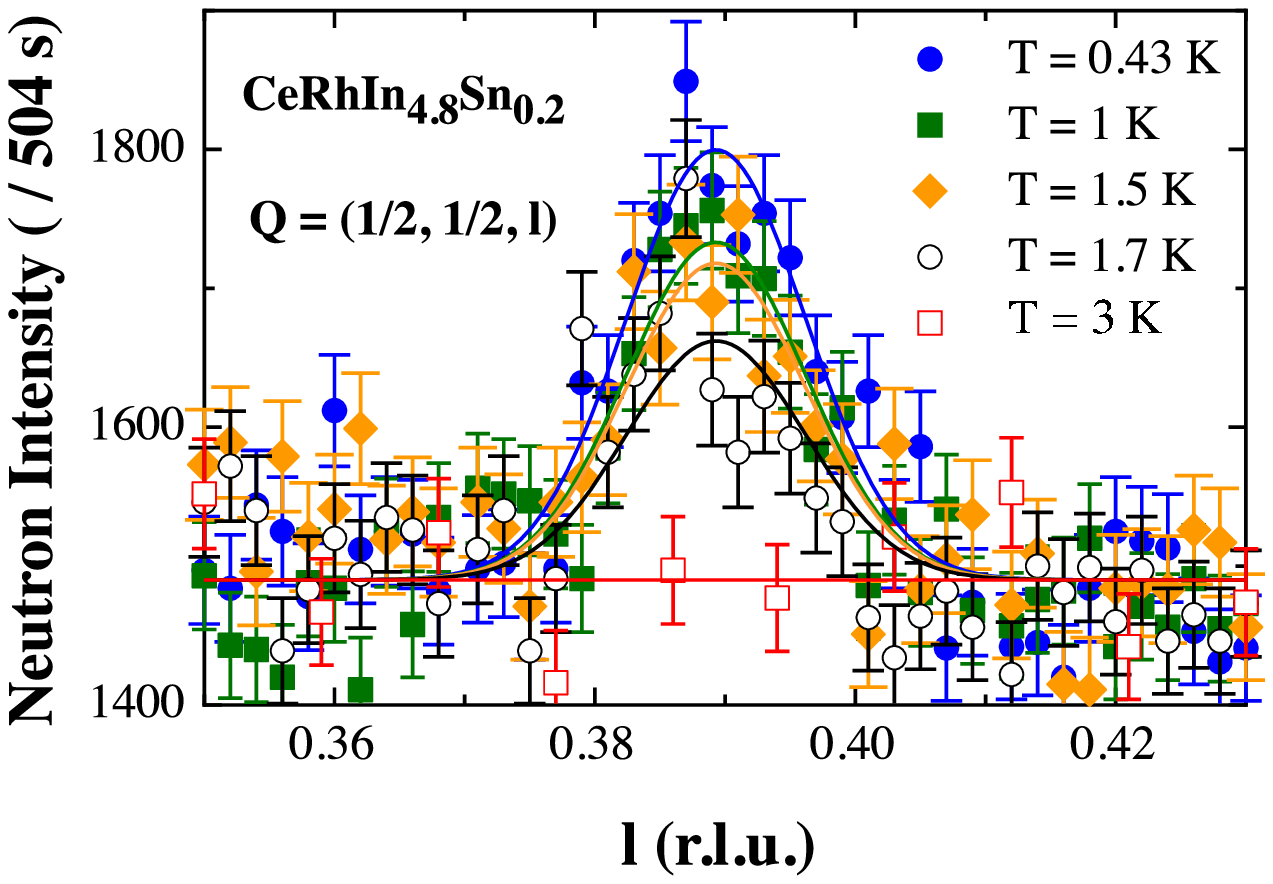}
\caption{Temperature dependence of a $\bf{Q}$-scan performed along [0,0,1] for $x$=0.20. Lines are Gaussian fits.}
\end{figure}

\begin{figure}
\centering
\includegraphics[width=8cm]{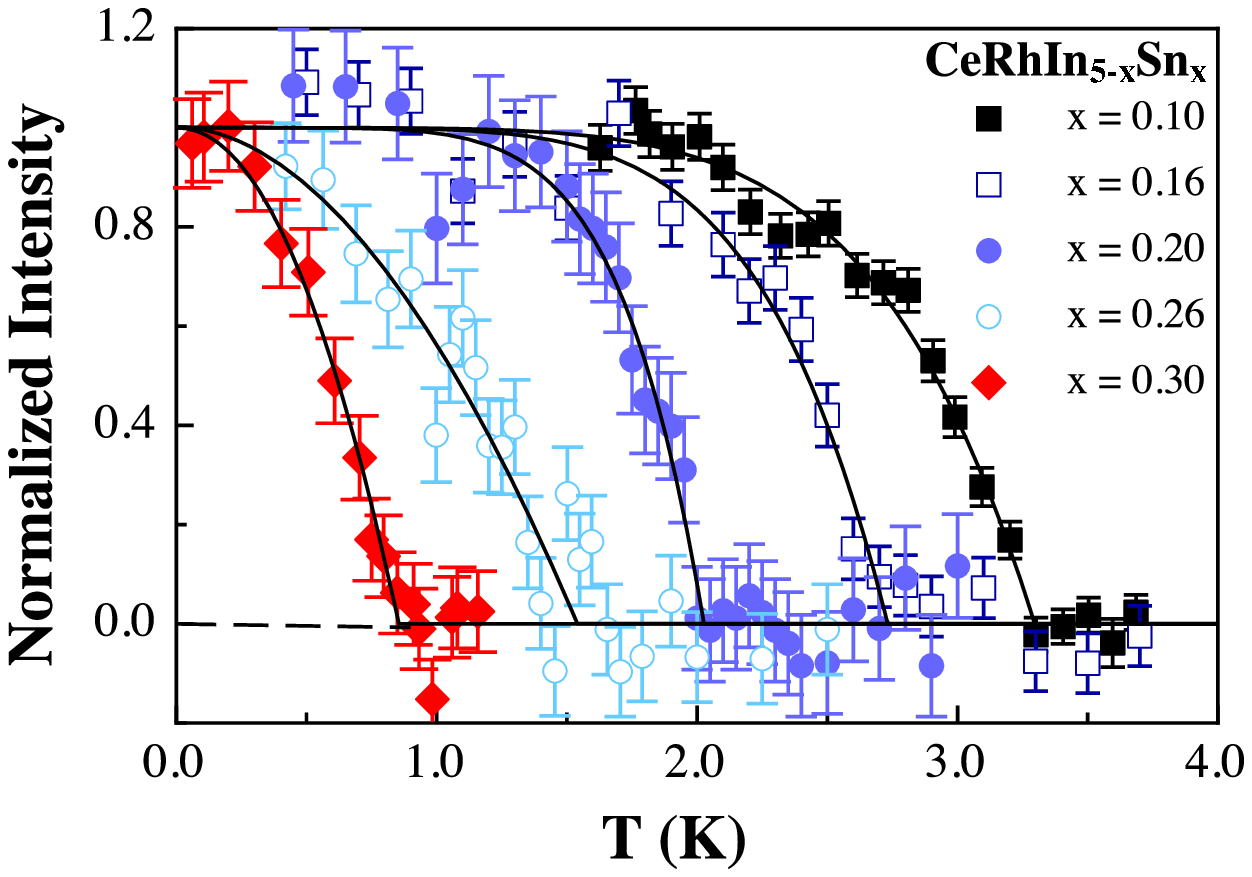}
\caption{Temperature dependence of the normalized magnetic intensities, for several $x$. Lines are phenomenological fits as explained in the text.} 
\end{figure}

\begin{figure}
\centering
\includegraphics[width=8cm]{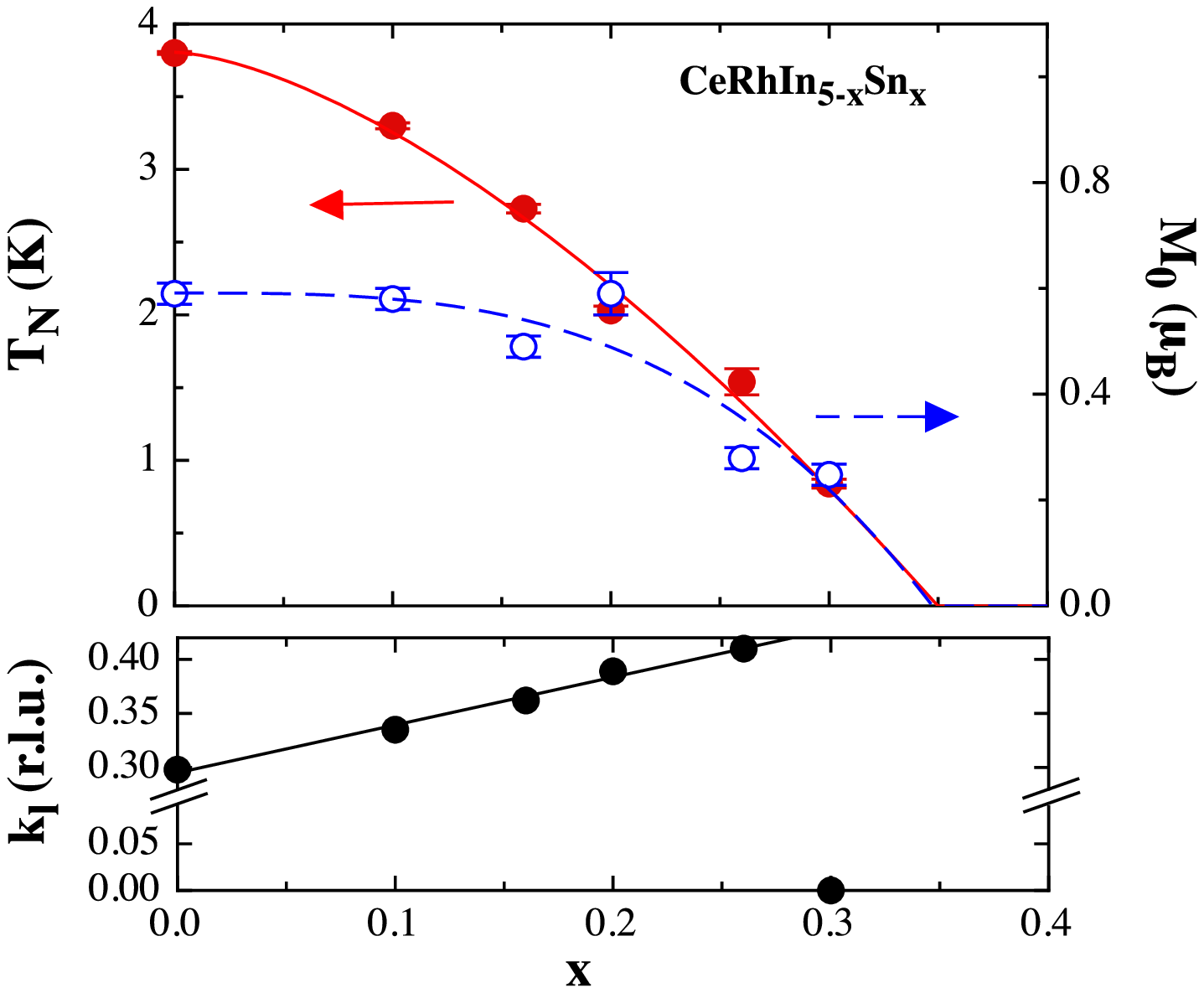}
\caption{N\'eel temperature, ordered magnetic moment and $c$-axis component of the propagation vector. Lines are guides for the eyes.}
\end{figure}


The main result of this study is the abrupt change of the propagation vector from incommensurate to commensurate in CeRhIn$_{5-x}$Sn$_{x}$ in the vicinity of its magnetic quantum critical point where superconductivity is expected to occur. It is to be specified that, up to now, firmly established bulk superconductivity superconductivity is never reported at zero pressure for any given $x$ but on applying 0.8 GPa starting from $x$=0.16 \cite{Mendonca} or 0.6 GPa starting from $x$=0.20 \cite{Knebel3}. These results suggest that further studies may evidence superconductivity at zero pressure in CeRhIn$_{5-x}$Sn$_{x}$ at higher $x$ near $x_c$. On the other hand, when $x$ increases, disorder increases and this may be detrimental to superconductivity. Having this in mind, one must nonetheless notice that similar changes of magnetic structure are already reported for several 1-1-5 related compounds for which superconductivity is firmly established.

A trend in the generic quantum critical ($x$, $P$, $T$) phase diagram of CeRhIn$_{5}$ related compounds is indeed the change from incommensurate to commensurate ordering associated with the appearance of superconductivity.
This is observed for Ir and Co doped CeRhIn$_{5}$ for which ordering with $\bf{k_{G}}$=(1/2, 1/2, 1/2) (G-type magnetic ordering) is reported either coexisting with or superseding the incommensurate ordering \cite{Christianson,Ohira,Yokoyama,Ohira2}. 
As concern CeRhIn$_{5}$ under pressure, NQR strongly suggests the same G-type commensurate ordering above 1.7 GPa \cite{Yashima}. This is not confirmed by neutron scattering experiments that were performed up to this pressure. Nonetheless, a switching from $k_{l}$ $\approx$ 0.30 to $k_{l}$ $\approx$ 0.40 is found at lower pressure in relation with superconductivity \cite{Raymond_P,Aso}.

All these data suggest that commensurate antiferromagnetism with $\bf{k_{G}}$=(1/2, 1/2, 1/2) is favorable for the formation of superconductivity in the quantum critical phase diagram of CeRhIn$_{5}$ related systems.
Strikingly, we also observe here a commensurate ordering in CeRhIn$_{5-x}$Sn$_{x}$ in the vicinity of $x_{c}$ but with $\bf{k_{C}}$=(1/2, 1/2, 0) instead of  $\bf{k_{G}}$ that would have been expected. 
This unachieved expectation was not  only built upon the aforementioned literature but also upon the puzzling fact that the extrapolation of $k_{l}(x)$ to $k_{l}$=1/2 occurs for $x$ $\approx$ $x_{c}$.
In addition, to our knowledge, a C-type magnetic ordering was never reported sofar for rare-earth based 1-1-5 systems. This propagation vetor is nonetheless the one of the magnetic order of the actinide based compound NpFeGa$_{5}$\cite{Honda} and of several rare-earth based compounds related to the 1-1-5 ones by different sequences of atomic stacking \cite{Cermak}.

As often pointed out, the Fermi surface topology is likely to play a key role for the determination of the magnetic ordering wavevector. This is specifically demonstrated for CeRhIn$_{5}$ by an {\it ab initio} calculation performed at $P$=0 that evidences a nesting of the Fermi surface for $k_{l}$=0.375, which is very close to the experimental value for the magnetic ordering wavevector $\bf{k_{IC}}$ \cite{Bjorkman}. dHvA experiments are very powerful to track the modification of the Fermi surface as a function of $P$ or $x$.
A change of Fermi surface from localized character to itinerant character occurs under pressure in CeRhIn$_{5}$ at around 2.3 GPa where the superconducting transition temperature is maximum \cite{Shishido}.
In a different way, Fermi surface reconstruction is also reported for Co substituted CeRhIn$_{5}$ when the magnetic structure switches from incommensurate to commensurate antiferromagnetism and when superconductivity occurs \cite{Goh}.
We can speculate that an abrupt modification of the Fermi surface occurs in CeRhIn$_{5-x}$Sn$_{x}$ between $x$=0.26 and $x$=0.30 and this drives the switching of the propagation vector.

The systems reviewed above realize a case where  $T_N$ is higher than $T_c$ and incommensurate magnetic ordering with $\bf{k_{IC}}$=(1/2, 1/2, $k_l$) seems to be detrimental to superconductivity.
The opposite situation ($T_{N}$ $\leq$ $T_{c}$) is also of great interest although experimental realization are scarce.
Recently we have shown that in Ce$_{0.95}$Nd$_{0.05}$CoIn$_{5}$, magnetic ordering with the incommensurate propagation vector $\bf{k_{Q}}$=(0.45, 0.45, 0.5) occurs \cite{Raymond_Nd}.
This is the same propagation vector as the one of the field induced antiferromagnetic phase of CeCoIn$_{5}$ starting from the pure $d$-wave superconducting state. Here again Fermi surface topology is believed to play a key role.
Since in both cases magnetic ordering occurs when superconductivity is established, it was suggested that $d$-wave superconductivity with nodes in the nesting area favors such an incommensurate order with in-plane incommensurability. 

Altogether these results suggest the possibility of collaborative effects between magnetism and superconductivity in the family of 1-1-5 compounds in relation with fine details of the Fermi surface.
While the involved mechanisms are not necessarily the same for all these systems, magnetism and superconductivity can either compete or collaborate in 1-1-5 systems. These two opposite situations are likely to originate from the position of  the nesting vector on the Fermi surface with respect to the superconducting order parameter nodes position. 

In summary, we evidence a switching of magnetic propagation vector from incommensurate with $\bf{k_{IC}}$=(1/2, 1/2, $k_l$) to commensurate with $\bf{k_{C}}$=(1/2, 1/2, 0) in CeRhIn$_{5-x}$Sn$_{x}$ in the proximity of its quantum critical point. Taking with caution the $P$-$x$ analogy, this would correspond to a region of the phase diagram where superconductivity arises.

We acknowledge K. Mony for help in sample preparation. Cerium was provided by the Materials Preparation Center, Ames Laboratory, US DOE Basic Energy Sciences, Ames, IA, USA, available from: www.mpc.ameslab.gov.


\begin{thebibliography}{99}


\bibitem{Uemura} Y. Uemura, Nature Materials 8 (2009) 253.
\bibitem{Sarrao} J. L. Sarrao and J. D. Thompson: J. Phys. Soc. Jpn. $\bf{76}$ (2007) 051013 and references therein.
\bibitem{Knebel} G. Knebel, D. Aoki and J. Flouquet, C.R. Physique $\bf{12}$, 542 (2011) and references therein. 
\bibitem{KnebelP} G. Knebel. D. Aoki, D. Braithwaite, B. Salce and J. Flouquet, Phys. Rev. B $\bf{74}$, 020501 (2006).
\bibitem{Park} T. Park, F. Ronning, H. Q. Yuan, M. B. Salamon, R. Movshovich, J. L. Sarrao and J. D. Thompson, Nature $\bf{440}$ (2006) 65.
\bibitem{Kenzelmann} M. Kenzelmann, Th. Str\"assle, C. Niedermayer, M. Sigrist, B. Padmanabhan, M. Zolliker, A. D. Bianchi, R. Movshovich, E. D. Bauer, J. L. Sarrao and J. D. Thompson, Science $\bf{321}$, 1652-1654 (2008) and references therein.
\bibitem{Yashima} M. Yashima, H. Mukuda, Y. Kitaoka, H. Shishido, R. Settai and Y. Onuki, Phys. Rev. B $\bf{79}$ (2009) 214528 and references therein.
\bibitem{Raymond_P} S. Raymond, G. Knebel, D. Aoki and J. Flouquet, Phys. Rev. B $\bf{77}$, 172502 (2008).
\bibitem{Aso} N. Aso, K. Ishii, H. Yoshizawa, T. Fujiwara, Y. Uwatoko, G.-F. Chen, N.K. Sato and K. Miyake, J. Phys. Soc. Japan $\bf{78}$, 073703 (2009).
\bibitem{Raymond_H} S. Raymond, E. Ressouche, G. Knebel, D. Aoki and J. Flouquet, J. Phys. Condens. Matter $\bf{19}$ (2007) 242204.
\bibitem{Bauer} E. Bauer, D. Mixson, F. Ronning, N. Hur, R. Movshovich, J. Thompson, J. Sarrao, M. Hundley, P. Tobash and S. Bobev: Physics B $\bf{378-80}$ (2006) 142.
\bibitem{Donath1} J. G. Donath, F. Steglich, E.D. Bauer, F. Ronning, J.L. Sarrao and P. Gegenwart, EPL $\bf{87}$, 57011 (2009).
\bibitem{Donath2} J. G. Donath, P. Gegenwart, F. Steglich, E.D. Bauer and J.L. Sarrao, Physica C $\bf{460-462}$, 661 (2007).
\bibitem{Mendonca} L. Mendon\c{c}a, T. Park, V. Sidorov, M. Nicklas, E.M. Bittar, R. Lora-Serrano, E.N. Hering, S.M. Ramos, M.B. Fontes, E. Baggio-Saitovich, H. Lee, J.L. Sarrao, J. D. Thompson and P.G. Pagliuso, Phys. Rev. Lett. $\bf{101}$ (2008) 017005.
\bibitem{Canfield} P.C. Canfield and Z. Fisk, Phil. Mag. B $\bf{65}$, 1117 (1992). 
\bibitem{Knebel2} G. Knebel, J. Buhot, D. Aoki, G. Lapertot, S. Raymond, E. Ressouche and J. Flouquet, J. Phys. Soc. Jpn. $\bf{80}$ (2011) SA001.
\bibitem{Rusz} J. Rusz, P.M. Oppeneer, N.J. Curro, R.R. Urbano, B.-L. Young, S. Lebegue, P.G. Pagliuso, L.D. Pham, E..D. Bauer, J.L. Sarrao and Z. Fisk, Phys. Rev. B $\bf{77}$, 245124 (2008).
\bibitem{Knebel3} G. Knebel, unpublished.
\bibitem{Christianson} A.D. Christianson, A. Llobet, W. Bao, J.S. Gardner, I.P. Swaison, J.W. Lynn, J.-M. Mignot, K. Prokes, P.G. Pagliuso, N.O. Moreno, J. L. Sarrao, J.D. Thomson and A.H. Lacerda, Phys. Rev. Lett. $\bf{95}$, 217002 (2005).
\bibitem{Ohira} S. Ohira-Kawamura, H. Shishido, A. Yoshida, R. Okazaki, H. Kawano-Furukawa, T. Shibauchi, H. Harima and Y. Matsuda, Phys. Rev. B $\bf{76}$, 132507 (2007).
\bibitem{Yokoyama} M. Yokoyama, N. Oyama, H. Maitsuka, S. Oinuma, I. Kawasaki, K. Tenya, M. Matsuura, K. Hirota and T.J. Sato, Phys. Rev. B $\bf{77}$, 224501 (2008).
\bibitem{Ohira2} S. Ohira-Kawamura, H. Kawano-Furukawa, H. Shishido, R. Okazaki, T. Shibauchi, H. Harima and Y, Matsuda, Phys. Status Solidi A $\bf{206}$ 1076 (2009).
\bibitem{Honda} F. Honda, N. Metoki, K. Kaneko, D. Aoki, Y. Homma, E. Yamamoto, Y. Shiokawa, Y. Onuki, E. Colineau, N. Bernhoeft and G. Lander, Physica B $\bf{359-361}$, 1147 (2005).
\bibitem{Cermak} P. \v{C}erm\'{a}k, P. Javorsk\'y, M. Kratochv\'\i lov\'{a}, Karel Pajskr, Milan Klicpera, Bachir Ouladdiaf, Marie-H\'el\`ene Lem\'ee-Cailleau, Juan Rodriguez-Carvajal and Martin Boehm, Phys. Rev. B $\bf{89}$ (2014) 184409 and references therein.
\bibitem{Bjorkman} T. Bjorkman, R. Lizarraga, F. Bultmark, O. Eriksson, J.M. Wills, A. Bergman, P.H. Andersson and L. Nordstrom, Phys. Rev. B $\bf{81}$, 094433 (2010).
\bibitem{Shishido} H. Shishido, R. Settai, H. Harima and Y. Onuki, J. Phys. Soc. Japan $\bf{74}$, 1103 (2005).
\bibitem{Goh} S. K. Goh, J. Paglione, M. Sutherland, E.C.T. O'Farell, C. Bergemann, T.A. Sayles and M.B. Maple, Phys. Rev. Lett. $\bf{101}$, 056402 (2008).
\bibitem{Raymond_Nd} S. Raymond, S.M. Ramos, D. Aoki, G. Knebel, V.P. Mineev and G. Lapertot, J. Phys. Soc. Japan $\bf{83}$, 013707 (2014).

\end{thebibliography}
\end{document}